\documentclass[conference]{IEEEtran} 
\ifCLASSINFOpdf
   \usepackage[pdftex]{graphicx}
\else
   \usepackage[dvips]{graphicx}
\fi
\usepackage[cmex10]{amsmath}
\usepackage{caption}
\usepackage{subcaption}

\usepackage{float}

\usepackage{url}

\usepackage[ruled,vlined]{algorithm2e}

\usepackage{comment}

\usepackage{amsthm}
\newtheorem*{remark}{Remark}
\newtheorem{insight}{Implementation Insight}

\makeatletter
\newenvironment{sublemma}[1]{%
  \def\subtheoremcounter{#1}%
  \refstepcounter{#1}%
  \protected@edef\theparentnumber{\csname the#1\endcsname}%
  \setcounter{parentnumber}{\value{#1}}%
  \setcounter{#1}{0}%
  \expandafter\def\csname the#1\endcsname{\theparentnumber.\Alph{#1}}%
  \ignorespaces
}{%
  \setcounter{\subtheoremcounter}{\value{parentnumber}}%
  \ignorespacesafterend
}
\makeatother
\newcounter{parentnumber}

\usepackage{xcolor}

\newtheorem{lemma}{Lemma}

\usepackage{graphicx,booktabs,array}
\usepackage{tabularx}

\usepackage{optidef}

\usepackage{algorithmic}

\IEEEoverridecommandlockouts 

\usepackage{amsfonts}

\usepackage{enumitem}

\usepackage[noadjust]{cite}

\begin{document}





\title{Using Voltage Phasor Control to Avoid Distribution Network Constraint Violations
\thanks{This work was supported by the U.S. Department of Energy, Award DE-EE0008008.}
} 



\author{
\IEEEauthorblockN{Keith Moffat, Alexandra von Meier}
\IEEEauthorblockA{Department of Electrical Engineering and Computer Science \\
University of California, Berkeley\\
Berkeley, USA\\
\{keithm, vonmeier\}@berkeley.edu}
}


\maketitle

\begin{abstract}
In this paper, we introduce Voltage Phasor Control (VPC), also known as Phasor Based Control, as a novel way of implementing Optimal Power Flow (OPF).  
Unlike conventional OPF, 
in which the power flow optimization broadcasts power injections, the VPC power flow optimization broadcasts voltage phasor setpoints to feedback controllers distributed throughout the network which respond to disturbances in real time. 
In this paper, we demonstrate that VPC can actively manage distributed energy resources to avoid voltage magnitude and line flow constraint violations in power distribution networks.
We provide sensitivities and bounds that quantify how the distributed voltage phasor feedback control reduces the effect of disturbances on the network voltages and line flows.
Using simulations, we compare the performance of VPC with the related but more conventional Voltage Magnitude Control (VMC). The sensitivities, bounds, and simulations provide implementation insights, which we highlight, for how distribution network operators can use VPC and VMC to avoid constraint violations.
\end{abstract}



\section{Introduction}

The proliferation of distributed solar generation and the electrification of transportation and heating end uses present a novel set of challenges for electricity distribution networks. 
For many distribution 
networks/grids/circuits,
these challenges include voltage magnitude and line flow constraint violations. 

As an alternative to upgrading distribution network infrastructure
to accommodate peak power demand/distributed generation, constraint violations can be mitigated by actively managing Distributed Energy Resources (DERs) such as batteries or controllable loads \cite{perez2017regulatory, horowitz2020techno}.
Both ``Distributed Energy Resource Management Systems'' (DERMS) and ``Active Distribution Networks'' (ADNs) \cite{gill2013dynamic, nick2014optimal, hidalgo2010review} actively manage DERs.

\begin{figure}[t]
\centering     
\includegraphics[scale=0.085]{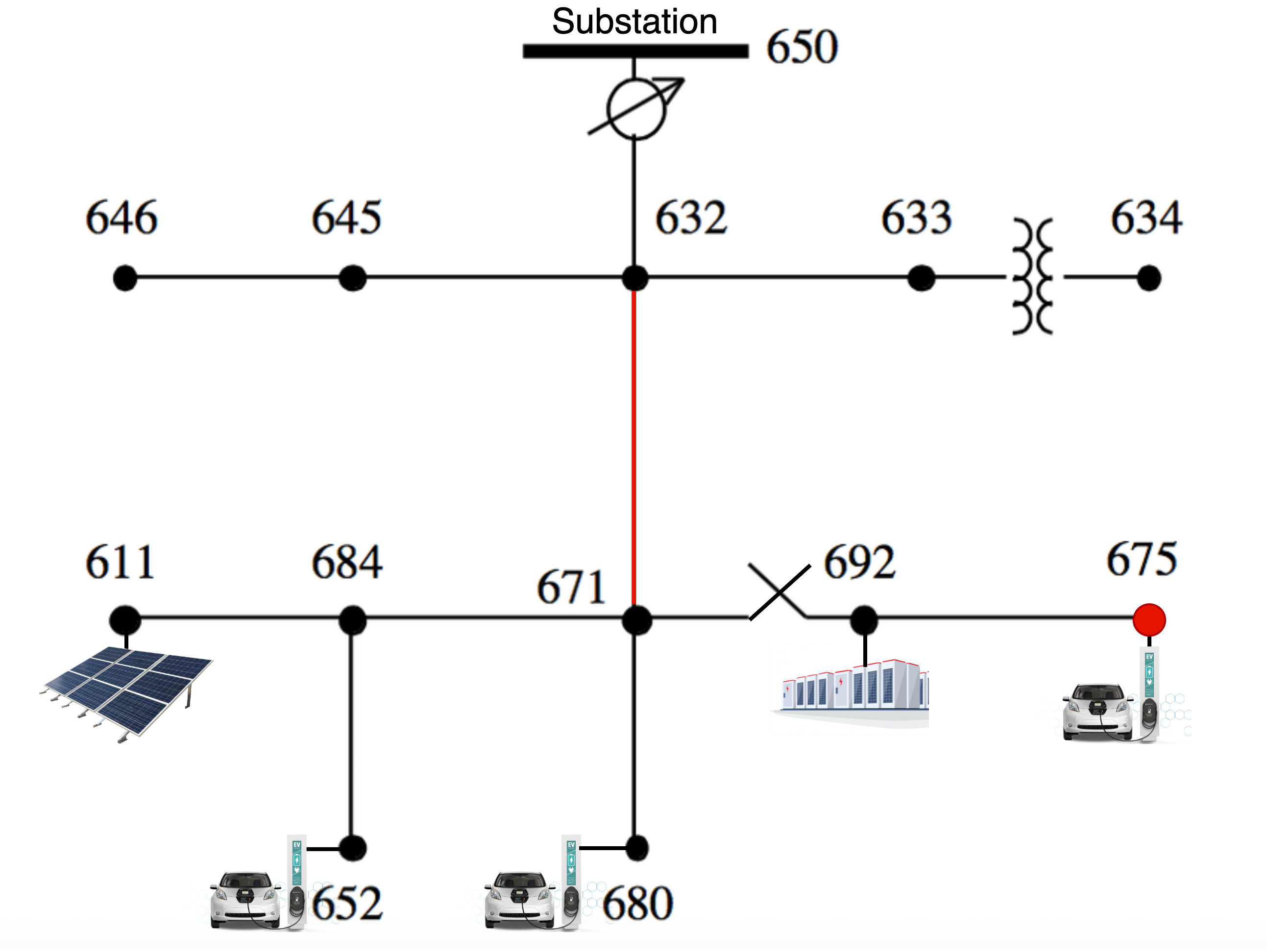}
\caption{\small The IEEE 13 Node Feeder with a line flow constraint violation on the line between nodes 632 and 671, and a voltage magnitude constraint violation at node 675. (The normally closed switch between nodes 671 and 692 is not relevant to this example.)}
\label{fig:13nodeConstraintViolation}
\end{figure}

Fig. \ref{fig:13nodeConstraintViolation} illustrates a simple example using the IEEE 13 Node Feeder. Suppose that large electric vehicle charging loads at nodes 652, 680, and 675 create a line flow/thermal constraint violation between nodes 632 and 671, and an undervoltage constraint violation at node 675. With a sufficiently large DER (e.g. a battery) at node 692, some combination of real and reactive power injections would be able to alleviate both the line flow and undervoltage constraint violations. 

In the transmission context, Optimal Power Flow (OPF) is used for economic dispatch, determining how much power each generator should produce subject to physical network constraints. 
OPF could also be used to actively manage DERs to enhance reliability and economic operation of distribution networks \cite{liu2021implementation}. 
However, the communication requirements of applying a power flow optimization \cite{bolognani2019need} may be problematic for distribution system applications. Because distribution networks benefit from less statistical aggregation of load than transmission networks, the ``disturbances,’’ or unanticipated changes in the real/reactive power injections on the network, occur at a faster rate. 
Thus, real and reactive power commands determined by an online distribution network OPF calculation may be outdated by the time they are implemented 
(e.g., if the cloud cover changes, affecting distributed solar generation). 

\begin{figure*}
    \centering
    \begin{subfigure}[t]{0.5\textwidth}
        \centering
        \includegraphics[width=0.85\textwidth]{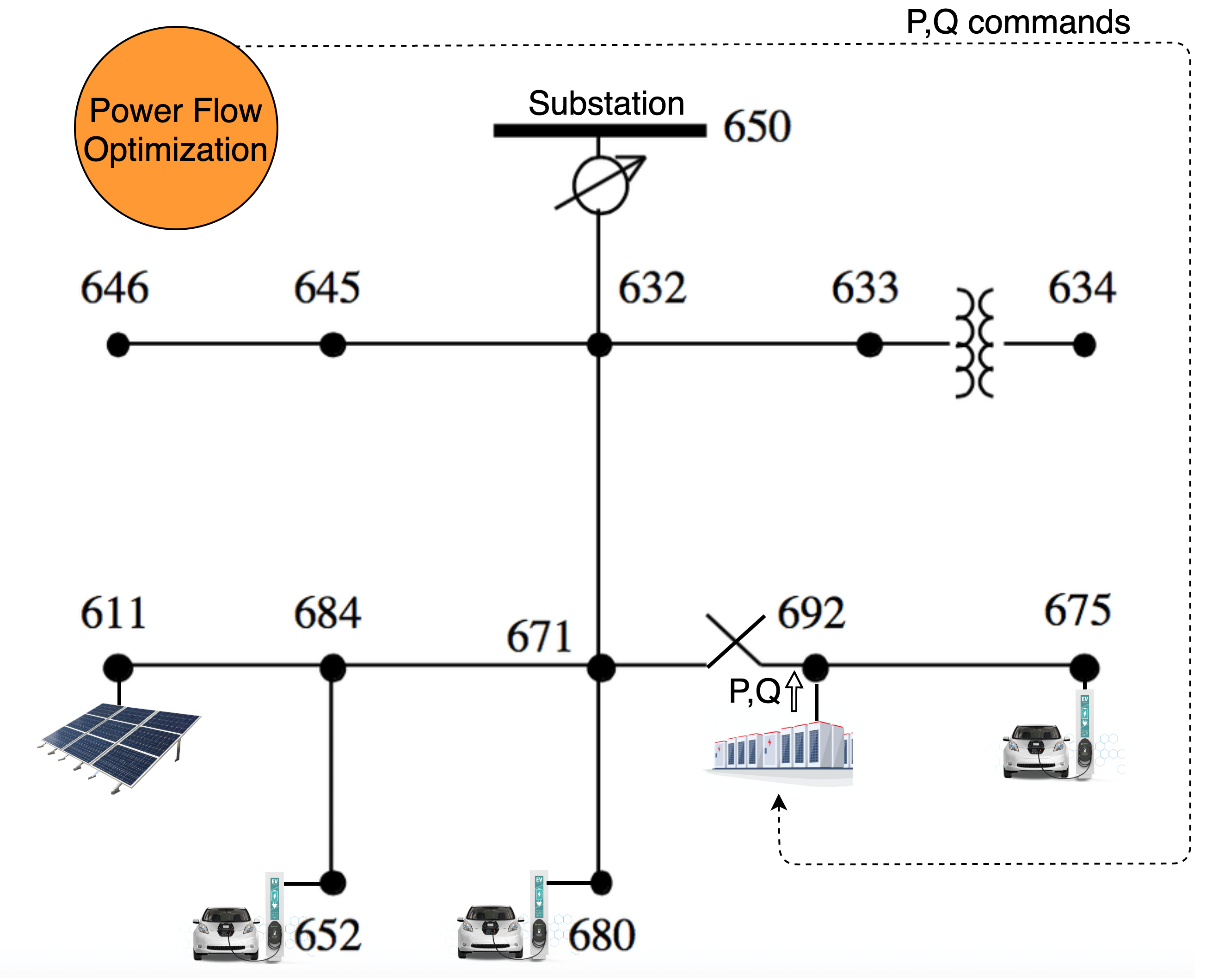}
        \caption{\small Centralized OPF implemented with direct power commands}
        \label{fig:13nodedirectPQcommand}
    \end{subfigure}%
    ~ 
    \begin{subfigure}[t]{0.5\textwidth}
        \centering
        \includegraphics[width=0.85\textwidth]{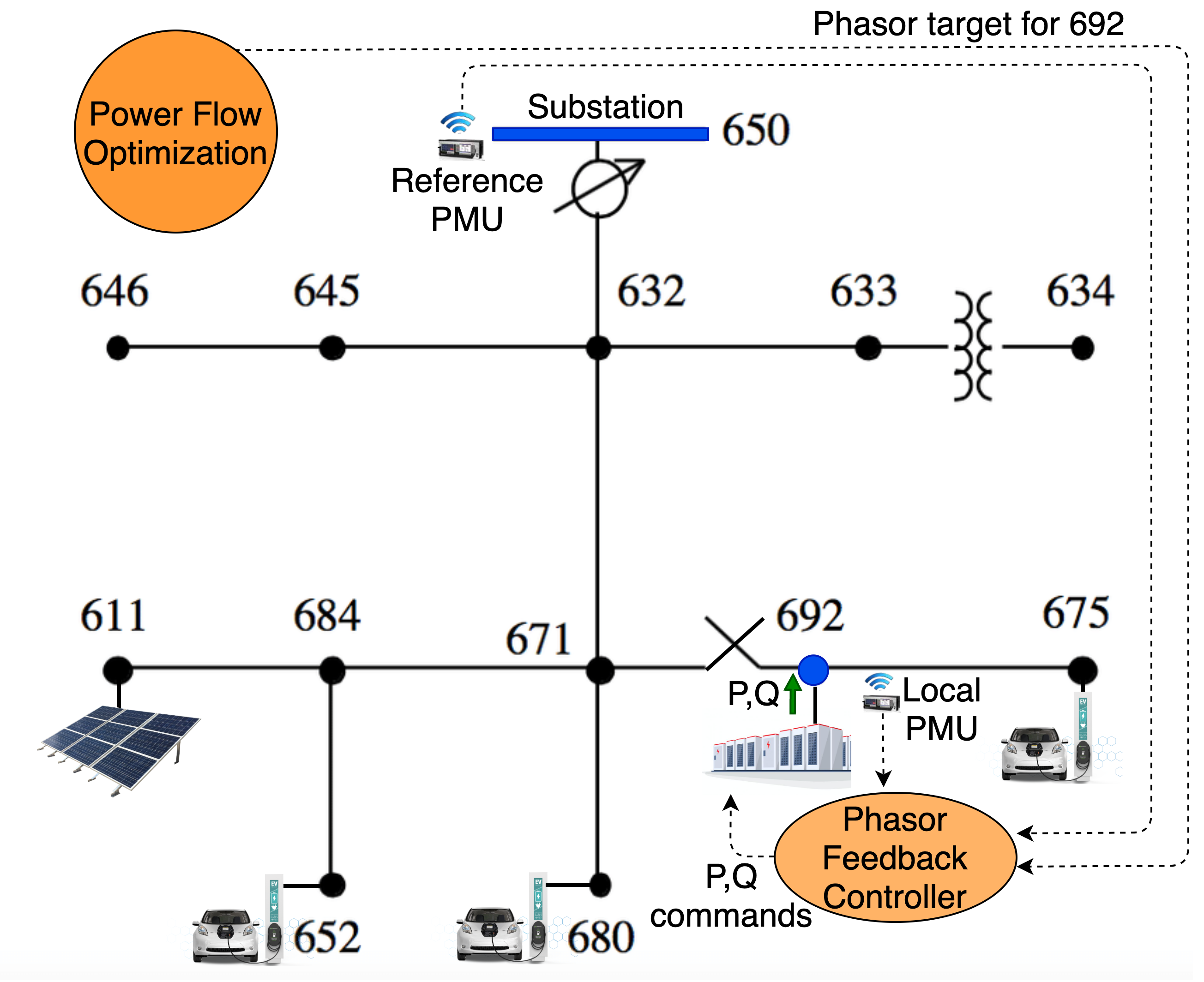}
        \caption{\small Voltage Phasor Control}
        \label{fig:13nodeVPC}
    \end{subfigure}
    \caption{\small Two different control methods for node 692 of the IEEE 13 Node Feeder}
    \label{fig:13nodeBoth}
\end{figure*}

On transmission networks, rather than implementing the centralized OPF solutions in open loop, the power generation decisions are adjusted in real time using distributed feedback loops including droop control, automatic voltage regulation, and Automatic Generation Control (AGC) based on the tie line flows \cite{cohn1956some, kumar2005recent, vrakopoulou2013probabilistic}. 
There are distributed feedback control methods for distribution networks as well, including Volt-Watt and Volt-VAR control \cite{horowitz2020techno, photovoltaics2018ieee ,farivar2013equilibrium, procopiou2019limitations, nacmanson2021advanced, hoke2019estimating, chathurangi2021comparative}, 
which define real or reactive power injections, respectively, as functions of voltage magnitude. 
Volt-Watt control can be combined with Volt-VAR control to avoid voltage violations, however the appropriate tradeoff between Volt-Watt and Volt-VAR responsibilities is situational 
\cite{rylander2016methods, bello2017optimal, yoshizawa2021voltage}.
\cite{baker2017network} provides an optimization-based approach to specifying the Volt-Watt and Volt-VAR curves that minimize voltage deviations and provide bounded-input-bounded-state stability. 
\cite{horowitz2020techno, procopiou2019limitations} show that, in some circumstances, Volt-VAR control can alleviate over-voltages on networks with high PV penetration, but can also increase line flow magnitudes.
Similarly, Fig. 9 in \cite{jain2020dynamic} demonstrates that Volt-VAR control can \emph{reduce} dynamic photovoltaic hosting capacity, if the line flow thermal constraints are binding. 

Voltage Phasor Control (VPC), also called Phasor Based Control \cite{vonMeier2020phasor}, is a novel paradigm for implementing OPF
which incorporates distributed feedback controllers, but in a different manner than AGC or Volt-Watt/Volt-VAR control.
With VPC,
the optimization broadcasts voltage phasor (magnitude and angle) setpoints, rather than real and reactive power setpoints, to participating nodes equipped with Phasor Measurement Units (PMUs).
VPC has been successfully demonstrated in Hardware-In-the-Loop simulations at Lawrence Berkeley Lab's FLEXLAB test facility \cite{moffat2021phasor, moffat2021linear}. 

In this paper, we compare the performance of VPC with Voltage Magnitude Control (VMC) \cite{molzahn2018towards, molzahn2019grid}, where the power flow optimization assigns voltage magnitudes and injection power factors to the distributed feedback controllers. When the injection power factor is 1, VMC corresponds to ``stiff'' Volt-Watt control, where the real power injection at each participating node is modulated to keep the local voltage magnitude equal to the assigned target. When the injection power factor is 0, VMC corresponds to stiff Volt-VAR control. 

Specifically, we analyze how effectively the real and reactive power injections under VPC's distributed feedback mitigate the adverse effects of disturbance power injections on the network voltages and upstream line flows.
By reducing the impact of disturbance injections, VPC feedback control allows the power flow optimization to be less overcautious than it would be with open loop power commands, and thus attain a more efficient/cost-effective operating point.




\begin{remark}
VPC is agnostic to the OPF implementation. It can be used with centralized OPF, as we assume in this paper, or with decentralized OPF.
\end{remark}


\section{Voltage Phasor Control}\label{sec:VPC}

An OPF solution contains both the optimal power injections and the corresponding voltage phasors. In standard OPF implementation, as demonstrated on the IEEE 13 Node Feeder in Fig. \ref{fig:13nodedirectPQcommand}, the OPF broadcasts power setpoints to the DER at node 692. In the Introduction, we describe this method as ``open loop'' because the DERs adhere to the power setpoints regardless of evolving grid conditions.

VPC is an alternative, ``closed loop'' method that broadcasts the OPF's voltage phasors to the participating Phasor Controlled Nodes (PCN).
Note, the OPF is only able to broadcast two set points because the power flow manifold permits two degrees of freedom for the state vector consisting of voltage magnitude $V$, voltage angle $\theta$, real power injection $P$, and reactive power injection $Q$ \cite{bolognani2015fast, hiskens2001exploring}. The table below displays how VPC and the standard $P Q$ and $P V$ control schemes relate to the power flow manifold.
\begin{table}[H]
    \centering
    \begin{tabularx}{.28\textwidth}{X c c c c }
       \toprule
        Control Type & $P$ & $Q$ & $V$ & $\theta$ \\
        \midrule
        $P Q$ Control & \textcolor{red}{x} & \textcolor{red}{x} &  &  \\
        $P V$ Control & \textcolor{red}{x} &  & \textcolor{red}{x} &  \\
        VPC &  &  & \textcolor{red}{x} & \textcolor{red}{x} \\
        \bottomrule
    \end{tabularx}
    \label{tab:PFmanifold}
\end{table}

Algorithm \ref{alg:VPC} describes how VPC is implemented. 
At each VPC node, a Phasor Feedback Controller (PFC) \cite{moffat2021localPVsens, moffat2021LQPC} adjusts the power injections in order to maintain its voltage phasor assignment despite evolving network disturbances. 
Fig. \ref{fig:13nodeVPC} demonstrates VPC with a single PCN at node 692.

\begin{algorithm}
\caption{\small \textbf{Voltage Phasor Control}}
\SetAlgoLined
\underline{\textbf{Optimization}}\\
At a slower cadence:\\
\begin{enumerate}
    \item Run OPF using load/generation predictions.
    \item Broadcast phasor targets to the PFCs.
    \item Gather measurements, adjust load predictions, repeat.
\end{enumerate}
\underline{\textbf{Distributed Phasor Feedback Controllers (PFCs)}}\\ 
Each PFC, at a faster cadence:\\
\begin{enumerate}
    \item Measure the local voltage phasor, receive the phasor reference angle from the substation.
    \item If a new phasor target has been received from the \\optimization, update phasor target.    
    \item Adjust the $P$ and $Q$ of subordinate DERs so that the local voltage phasor is closer to the phasor target, \\repeat.
\end{enumerate}
\label{alg:VPC}
\end{algorithm}

VPC requires synchronized phasor measurements from PMUs, rather than standard voltage magnitude measurements. Furthermore, the phasor measurements require a reference node that defines the ``0'' angle. For distribution networks, the substation is the natural choice for the phasor reference node as it behaves approximately like a slack bus, supplying the net power imbalance incurred by the load and generation at all of the other nodes on the network.

VPC is well-suited for distribution networks with volatile disturbances because the PFCs make power adjustment decisions in a distributed manner. 
Each PFC requires only the local phasor measurement and the reference phasor angle measurement, which is constantly broadcast to all of the PFCs, and therefore is able to immediately adjust the power injections of its subordinate DERs without waiting to hear from the OPF computation. 

\begin{remark}
While we consider power injection disturbances in this paper, VPC is applicable to other applications including switch closing, phase balancing, and responding to line outages on mesh networks. 
\end{remark}


\section{VPC and Voltage Disturbance Sensitivity}

\begin{figure}
\centering     
\includegraphics[scale=.12]{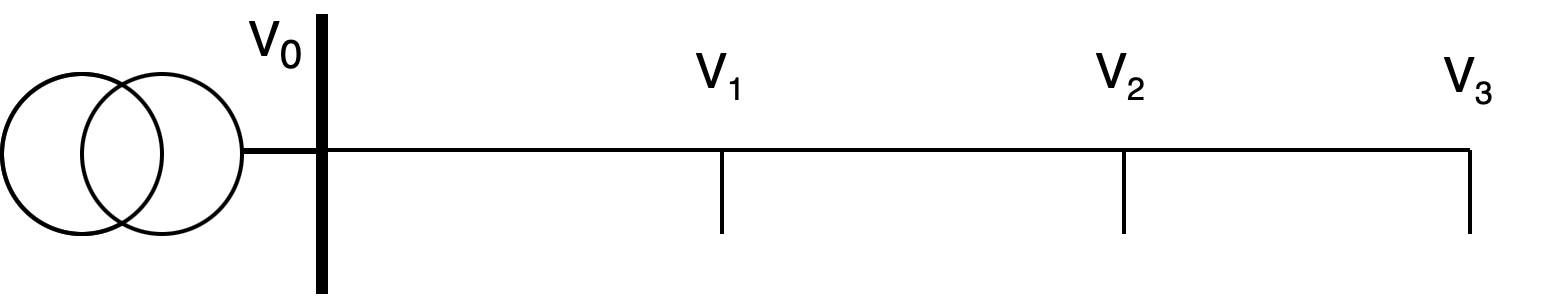}
\caption{\small Four node circuit}
\label{fig:4NodeNoMarkings}
\end{figure}

To clearly demonstrate the effect that disturbance injections have on the voltages on the network, we use the simplest possible circuit---the four node circuit in Fig. \ref{fig:4NodeNoMarkings}. Node 0 is the substation, which we model as an infinite/slack bus with voltage 1 per unit (p.u.). 
We consider scenarios in which the circuit has one PCN, one disturbance injection node which is perturbed by $i_\textnormal{dist}$, and one ``Node of Concern’’ (NC)---the node at which we are concerned about voltage magnitude constraint violations. 
We consider the four PCN/NC/disturbance configurations in Table \ref{tab:Vsens},\footnote{We do not include the scenarios which correspond to the circuits in rows 3 and 4 with the NC and disturbance nodes switched because these scenarios are similar to the circuits in rows 3 and 4.}
and make the assumption that both the pre-disturbance injections and the disturbance injections are constant-current to simplify the circuit analysis.\footnote{We choose to describe the disturbance as a current injection (rather than a power injection) because it makes the NC voltage linear with respect to the disturbance current injection, even when VPC holds the PCN constant. 
Note, this linearity is a convenient property of VPC only. It does not apply to circuits with VMC nodes.
Linear equations simplify the sensitivity analysis because they are analytic and therefore the complex derivative is defined.}
Columns 3 and 4 display the sensitivities of the NC voltage to the disturbance current injection when the system is run in open loop and when VPC is applied, respectively.  
The circuit equations and derivations of the first two sensitivities for Table \ref{tab:Vsens} are in Appendix \ref{app:Vsens}.

\begin{table*}
    \centering
    \begin{tabularx}{.73\textwidth}{cccc}
       \toprule
        Row & Configuration & Open Loop Sensitivity & Sensitivity With VPC \\
        \midrule
        1 &  \raisebox{-.5\height}{\includegraphics[scale=.1]{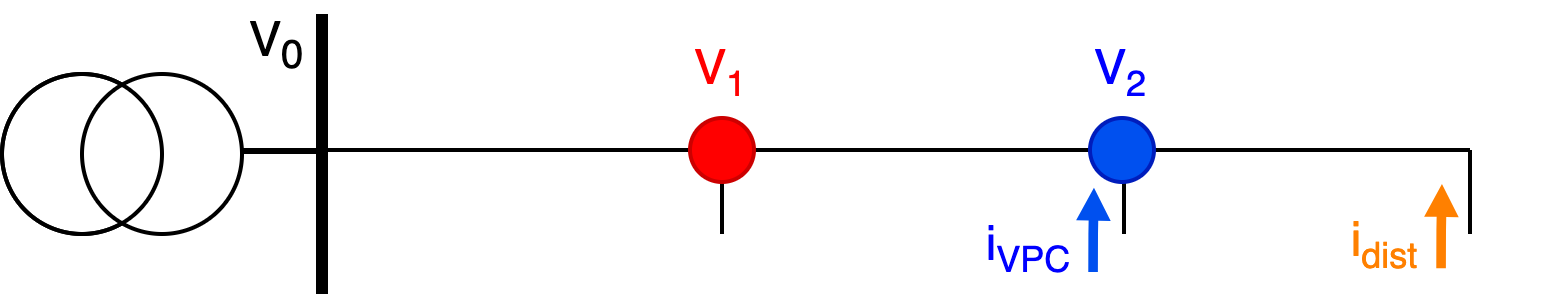}} & $\frac{\partial v_1}{\partial i_3} = z_{01}$ & $\frac{\partial v_1}{\partial i_3} = 0$  \\
        2 &  \raisebox{-.5\height}{\includegraphics[scale=.1]{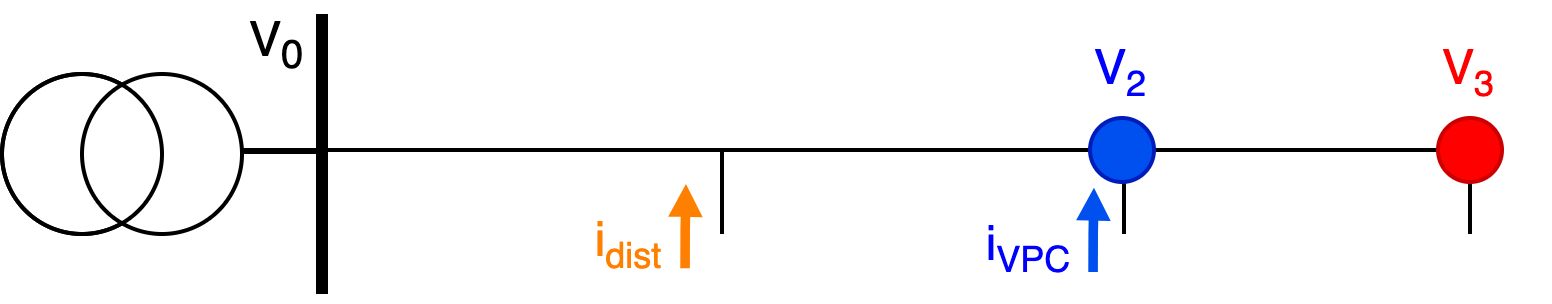}} & $\frac{\partial v_3}{\partial i_1} = z_{01}$ & $\frac{\partial v_3}{\partial i_1} = 0$  \\
        3 &  \raisebox{-.5\height}{\includegraphics[scale=.1]{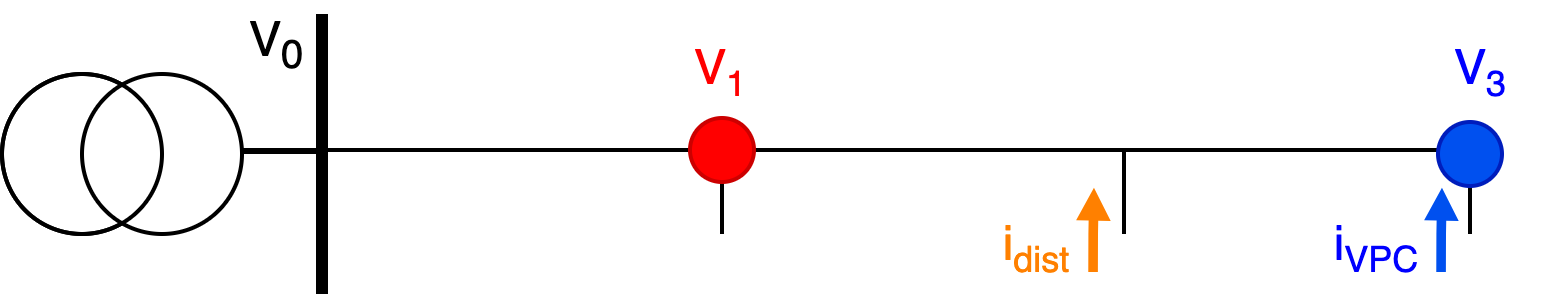}} & $\frac{\partial v_1}{\partial i_2} = z_{01}$ & $\frac{\partial v_1}{\partial i_2} = \frac{z_{01}}{z_{23} + z_{12} + z_{23}}z_{01}$  \\
        4 & \raisebox{-.5\height}{\includegraphics[scale=.1]{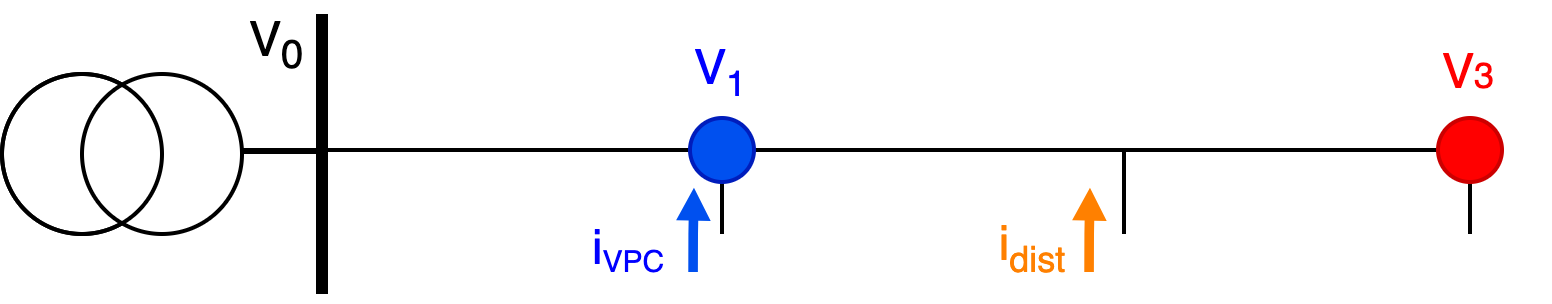}} & $\frac{\partial v_3}{\partial i_2} = z_{01} + z_{12}$ & $\frac{\partial v_3}{\partial i_2} = z_{12}$ \\ 
        \bottomrule
    \end{tabularx}
    \caption{\small Voltage disturbance sensitivities. The phasor controlled node (PCN) is blue and the node of concern (NC) is red.}    
    \label{tab:Vsens}
\end{table*}

For the configurations in rows 1 and 2,\footnote{The circuits in Table \ref{tab:Vsens} can be thought of as an equivalent circuit models for larger circuits. 
For example, the configuration in row 2 corresponds to the IEEE 13 Node Feeder circumstance in which node 675 is the NC, there is a PCN at node 692 and there is a disturbance injection at node 611.} for which the disturbance is on the other side of the PCN than the NC, the disturbance has no effect on the voltage at the NC. That is, the PCN injection fully cancels the effect of the disturbance injection on the NC voltage. 
In the bottom two configurations the NC is on the same side of the VPC node as the disturbance. In these circumstances, for most $z_{01}$ and $z_{12}$ values, the PCN injection reduces the effect of the disturbance on the voltage, but does not eliminate it. 


\begin{insight}
When concerned with voltage magnitude(s) of a node or set of nodes, a PCN should be placed close to the node(s) of concern, preferably between the node(s) of concern and the disturbance node(s).
\end{insight}

    
    

    
    


\section{VPC and Upstream Line Flow Disturbance Sensitivity}

To clearly demonstrate the effect that disturbance injections have on upstream line flows for radial circuits, we use the same four node circuit in Fig. \ref{fig:4NodeNoMarkings} and again make the assumption that the injections are constant-current.
With this assumption, VPC only affects upstream line flows. 
In practice, VPC affects downstream line flows as well, due to the voltage dependencies of the downstream loads and generators. However these effects are second order and may not be significant. 

Similar to the voltage sensitivities, we define the ``Line of Concern’’ (LC) for a given configuration as the line at which we are concerned about line flow constraint violations. 
We consider the three PCN/LC/disturbance configurations in Table \ref{tab:Lfsens}.\footnote{The circuits in Table \ref{tab:Lfsens} can also be thought of as equivalent circuit models for larger circuits. 
For example, the configuration in row 3 corresponds to the circumstance in which the line between nodes 671 and 632 is the LC, there is a PCN at 692, and there are disturbance injections at 652 and 680.} 
The circuit equations and derivations for Table \ref{tab:Lfsens} are in Appendix \ref{app:LFsens}.



\begin{table*}
    \centering
    \begin{tabularx}{.63\textwidth}{cccc}
       \toprule
        Row & Configuration & Open Loop Sensitivity & Sensitivity With VPC \\
        \midrule
        1 &  \raisebox{-.5\height}{\includegraphics[scale=.1]{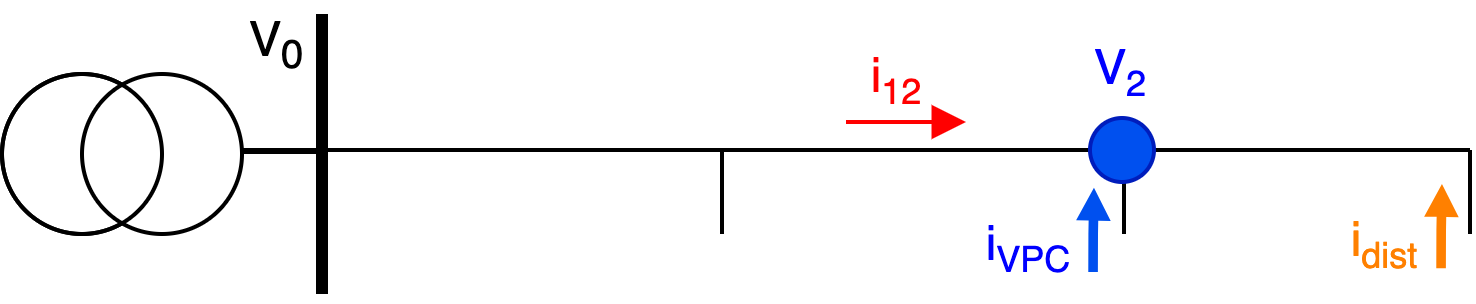}} & $\frac{\partial i_{01}}{\partial i_3} = -1$ & $\frac{\partial i_{01}}{\partial i_3} = 0$  \\
        2 &  \raisebox{-.5\height}{\includegraphics[scale=.1]{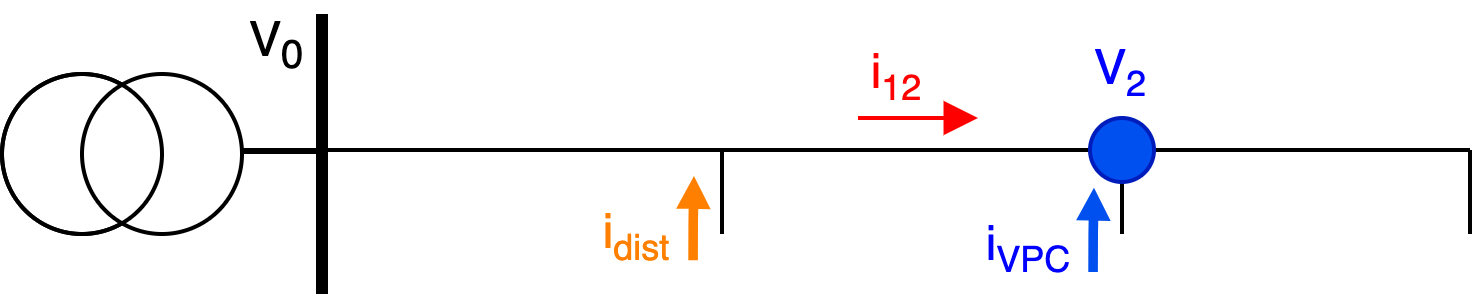}} & $\frac{\partial i_{12}}{\partial i_1} = 0$ & $\frac{\partial i_{12}}{\partial i_1} = \frac{z_{01}}{z_{01} + z_{12}}$  \\           
        3 &  \raisebox{-.5\height}{\includegraphics[scale=.1]{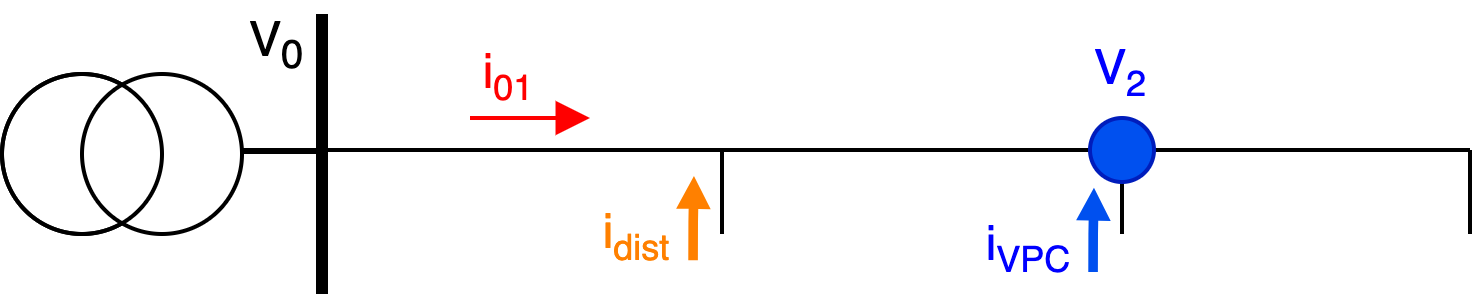}} & $\frac{\partial i_{01}}{\partial i_1} = -1$ & $\frac{\partial i_{01}}{\partial i_1} = -\frac{z_{12}}{z_{01} + z_{12}}$  \\
        \bottomrule
    \end{tabularx}
    \caption{\small Line flow disturbance sensitivities. The phasor controlled node (PCN) is blue and the line flow of concern (LC) is red.} 
    \label{tab:Lfsens}
\end{table*}

In the circuit in row 1 of Table \ref{tab:Lfsens}, in which the disturbance is downstream of the PCN, the PCN completely shields the upstream line flows from downstream disturbance injections. 
In the circuit in row 2, in which the disturbance is upstream of the PCN and the LC, the VPC can either increase or decrease $|i_{12}|$ depending on the 
circumstance.
In the circuit in row 3, in which the disturbance is upstream of the PCN but downstream of the LC, the disturbance affects $i_{01}$ but
VPC reduces the likelihood of $i_\textnormal{dist}$ producing a $|i_{01}|$ constraint violation.


\subsection{Bounds on changes in the upstream current magnitude}\label{subsec:VPCupstreamBounds}

Regarding line flow constraint violations, we are interested in the disturbance-sensitivity of current magnitudes.
Unfortunately, generally 
\begin{gather*}
    |\frac{\partial i_{01}}{\partial i_1}| \neq \frac{\partial |i_{01}|}{\partial i_1}.
\end{gather*}
Furthermore, the magnitude operator is not analytic, and therefore we cannot use a standard derivative. 

Thus, we are left with conservative bounds on the changes of $|i_{01}|$. To articulate these bounds, we define the following variables for $i_{01}$ and corresponding quantities for $i_{12}$:
\begin{itemize}
    \item $\boldsymbol{\Delta i_1}$: The disturbance injection current at node 1.
    \item $\boldsymbol{   }$: The $i_{01}$ current before $\Delta i_1$ is added to the injection at node 1.
    \item $\boldsymbol{i^\textnormal{\textbf{ol}}_{01}}$: The $i_{01}$ current after $\Delta i_1$ is added when the system is run open loop (ol).
    \item $\boldsymbol{i_{01}^\textnormal{\textbf{VPC}}}$: The $i_{01}$ current after $\Delta i_1$ is added when VPC holds node 2's voltage phasor constant.
\end{itemize}
For clarity, Lemmas \ref{lemma:DeltaCurr} and \ref{lemma:CurrDiff} 
assume the pre-disturbance injections are constant-current and the disturbance is defined as a current injection as well.

\begin{sublemma}{lemma}\label{lemma:DeltaCurr}
\begin{lemma}\label{lemma:DeltaCurr01}
$|i_{01}^\textnormal{VPC}| - |i^\textnormal{bef}_{01}| \leq |\Delta i_1| |\frac{z_{12}}{z_{01} + z_{12}}|$.
\end{lemma}

\begin{lemma}\label{lemma:DeltaCurr12}
$|i_{12}^\textnormal{VPC}| - |i^\textnormal{bef}_{12}| \leq |\Delta i_1| |\frac{z_{01}}{z_{01} + z_{12}}|$.
\end{lemma} 
\end{sublemma}
\noindent Lemma \ref{lemma:DeltaCurr}, proved in Appendix \ref{app:LemmaDeltaCurr}, bounds the changes in the upstream current magnitudes in terms of the disturbance magnitude $|\Delta i_1|$ and the line impedances.



While Lemma \ref{lemma:DeltaCurr} bounds the changes in the upstream line flows, it does not state whether VPC \emph{always} helps or hurts when a disturbance increases the upstream line flows. For the circuit in row 2 of Table \ref{tab:Lfsens}, the effect of VPC depends on the directions of $i^\textnormal{bef}_{12}$ and $\Delta i_{1}$ in the complex plane, as well as the line impedances. For the circuit in row 3, however, we can bound the difference between $|i_{01}^\textnormal{VPC}|^2$ and $|i_{01}^\textnormal{ol}|^2$ for all disturbances that increase $|i_{01}^\textnormal{ol}|$.
\begin{lemma}
For the circuit in row 3 of Table \ref{tab:Lfsens}, if $z_{01}$ and $z_{12}$ have the same X/R ratio, and if $\Delta i_1$ increases $|i_{01}|$, then
\begin{gather*}
    |i_{01}^\textnormal{ol}|^2 - |i_{01}^\textnormal{VPC}|^2 > (1 - a)|\Delta i_1|^2 > 0,\\
    \textnormal{where } a = \frac{z_{12}}{z_{01} + z_{12}} \in [0,1] \subset \mathbb{R}.
\end{gather*}
\label{lemma:CurrDiff}
\end{lemma}
\noindent Lemma \ref{lemma:CurrDiff}, proved in Appendix \ref{app:Lemma:CurrDiff}, states that if the lines between nodes 0 and 1 and nodes 1 and 2 have the same X/R ratio, then the VPC at node 2 reduces the risk that $\Delta i_1$ will create an $|i_{01}|$ constraint violation, regardless of the pre-disturbance $i_{01}$, $i_1$, and $\hat{v}_2$ values.
Furthermore, the intuition derived from the proof of Lemma \ref{lemma:CurrDiff} is that if the X/R ratios for $z_{01}$ and $z_{12}$ are not pathologically different, then VPC reduces the risk that $\Delta i_1$ will create an $|i_{01}|$ constraint violation.


\subsection{Assessing the conservativeness of the upstream line flow bounds}

The bounds in Lemmas \ref{lemma:DeltaCurr} and \ref{lemma:CurrDiff} are conservative.
To assess the conservativeness of the bounds, we ran a test on the circuit in row 3 of Table \ref{tab:Lfsens} which compared the Lemma bounds with the observed change in $|i_{01}|$ for a range of disturbance injections at node 1. 
We set the line impedances $z_{01} = z_{12} = 0.5 + 0.5j$. 
The current magnitudes were normalized so that the pre-disturbance $|i_{01}| = 1$.
The pre-disturbance voltage magnitudes at nodes 1 and 2 were 0.97 and 0.95 respectively.
Disturbance injections with a range of power factors, from 0 to 1 lagging (extracting reactive power), and from 1 to 0.7 leading (injecting reactive power), were injected at node 1.
The VPC at node 2 adjusted its injection in order to maintain its assigned voltage phasor. 

\begin{figure}
\centering     
\includegraphics[scale=.22]{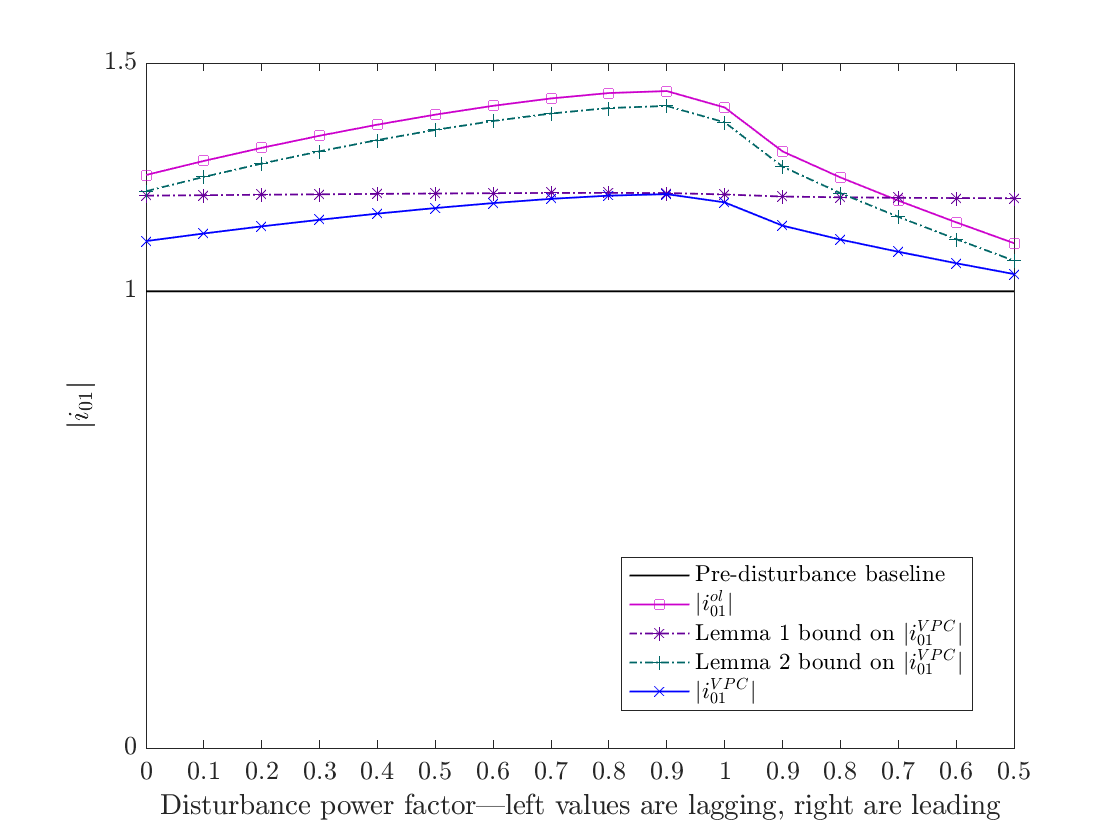}
\caption{\small Comparison of the bounds in Lemmas \ref{lemma:DeltaCurr} and \ref{lemma:CurrDiff} for the circuit in row 3 of Table \ref{tab:Lfsens} for disturbances with a range of power factors.}
\label{fig:boundCompare}
\end{figure}

Fig. \ref{fig:boundCompare} plots $|i_{01}|$ vs. the power factor of the disturbance at node 1 and demonstrates that the Lemma \ref{lemma:DeltaCurr01} bound ranges between conservative and tight. For this example, the Lemma \ref{lemma:DeltaCurr01} bound does not assert that VPC reduces the increase in $|i_{01}|$ when the disturbance has a leading power factor of 0.7 or less. 
Fig. \ref{fig:boundCompare} also demonstrates that, while the Lemma \ref{lemma:CurrDiff} bound is generally quite conservative, the Lemma \ref{lemma:CurrDiff} bound does accurately assert that VPC reduces the increase in the upstream line flow current magnitudes for all disturbance power factors.

 

\begin{insight}
When concerned with a line flow or set of line flows, a PCN should be placed downstream, but as close as possible, to the line flow(s) of concern.
\end{insight}

\section{Voltage Magnitude Control}

As stated in Section \ref{sec:VPC}, the power flow manifold permits two degrees of freedom. 
Thus, a voltage magnitude command for a given node must be accompanied by an additional state command. 
We choose the Adjustment Power Factor (APF) of the feedback control as the second state command, which includes stiff Volt-Watt and Volt-VAR control as the APF = 1 and APF = 0 special cases, respectively, and implement VMC with Algorithm \ref{alg:VMC}. 


\begin{algorithm}
\caption{\small \textbf{Voltage Magnitude Control}}
\SetAlgoLined
\underline{\textbf{Optimization}}\\
At a slower cadence,\\
\begin{enumerate}
    \item Run OPF using load/generation predictions.
    \item Broadcast magnitude targets and APFs to the MFCs.
    \item Gather measurements, adjust load predictions, repeat.
\end{enumerate}
\underline{\textbf{Distributed Magnitude Feedback Controllers (MFCs)}}\\ 
Each MFC, at a faster cadence,\\
\begin{enumerate}
    \item Measure the local voltage magnitude.
    \item If a new magnitude target and/or APF has been \\received, update the magnitude target and/or APF. 
    \item Adjust the $P$ and $Q$ of subordinate DERs in way that maintains the APF and brings the local voltage magnitude closer to the magnitude target, repeat.
\end{enumerate}
\label{alg:VMC}
\end{algorithm}


While there always exist real and reactive power adjustments at a given node that produce any voltage phasor assignment, there does not always exist real and reactive power adjustments with a given APF that produce any voltage magnitude assignment.
That is, a function can be defined from an arbitrary domain of voltage phasors to the appropriate codomain of power adjustment using Ohm's Law. 
However not all APF/voltage magnitude target combinations can be mapped to power injections on the power flow manifold \cite{hiskens2001exploring}. 
For example, consider a predominantly reactive network in which an MCN is assigned APF $= 1$ (stiff Volt-Watt control). Adjusting the real power injection will have limited effect on the voltage magnitude, thus many voltage magnitude assignments are infeasible.


\begin{insight}
Care must be taken when selecting APFs for VMC. Assigning an infeasible voltage magnitude target/APF will produce voltage instabilities.
\end{insight}




We observed in simulations that VMC is effective in reducing the impact of disturbances on voltage magnitudes in the neighborhood of a given MCN node, as expected.
As we highlight in the next section, while VMC usually helps upstream line flow constraint violations, it generally does so less effectively than VPC and, in some cases, exacerbates the increase in upstream line flow caused by the disturbance.

\section{Comparing VPC and VMC for Upstream line flow Constraint Violations}

To demonstrate how VPC and VMC reject the effect of disturbances on upstream line flows, we ran 
\ifdefined\fullresults
three 
\else
two
\fi
sets of simulations similar to the set that created Fig. \ref{fig:boundCompare}. We used the circuit in row 3 of Table \ref{tab:Lfsens} and again set $z_{01} = z_{12} = 0.5 + 0.5j$. 
\ifdefined\fullresults
On the x--axis, we swept through a set of power factors for the node 1 disturbance. 
We plotted the $|i_{01}|$ baseline and $|i_{01}|$ with no control, VPC at node 2, or VMC at node 2. 
\fi



\ifdefined\fullresults
\subsection{Simulation set I: non-nominal baseline, high voltage example}
\else
\subsection{Simulation set I: non-nominal baseline}
\fi 

\begin{figure}
\centering     
\includegraphics[scale=.22]{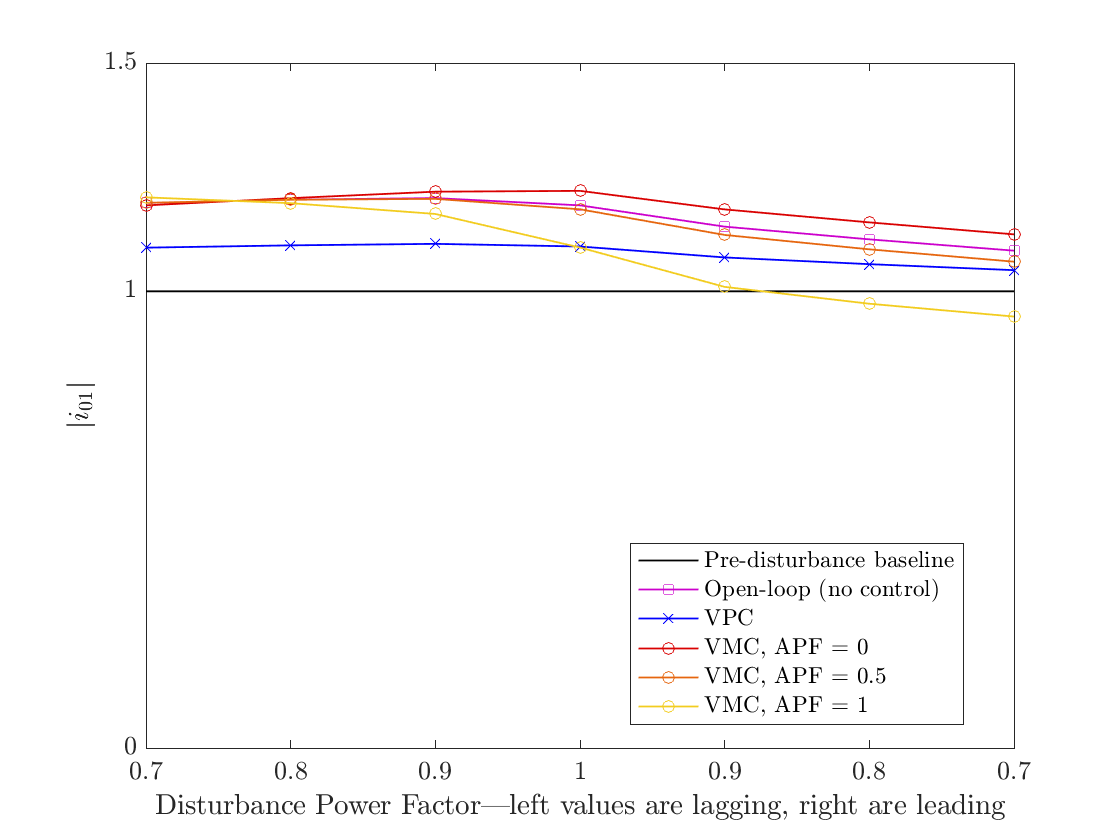}
\caption{\small Line flows for the circuit in row 3 of Table \ref{tab:Lfsens} with a $|v_2| = 1.05$ non-nominal pre-disturbance baseline.}
\label{fig:highVnonNominalLineflows}
\end{figure}

\begin{figure}
\centering     
\includegraphics[scale=.22]{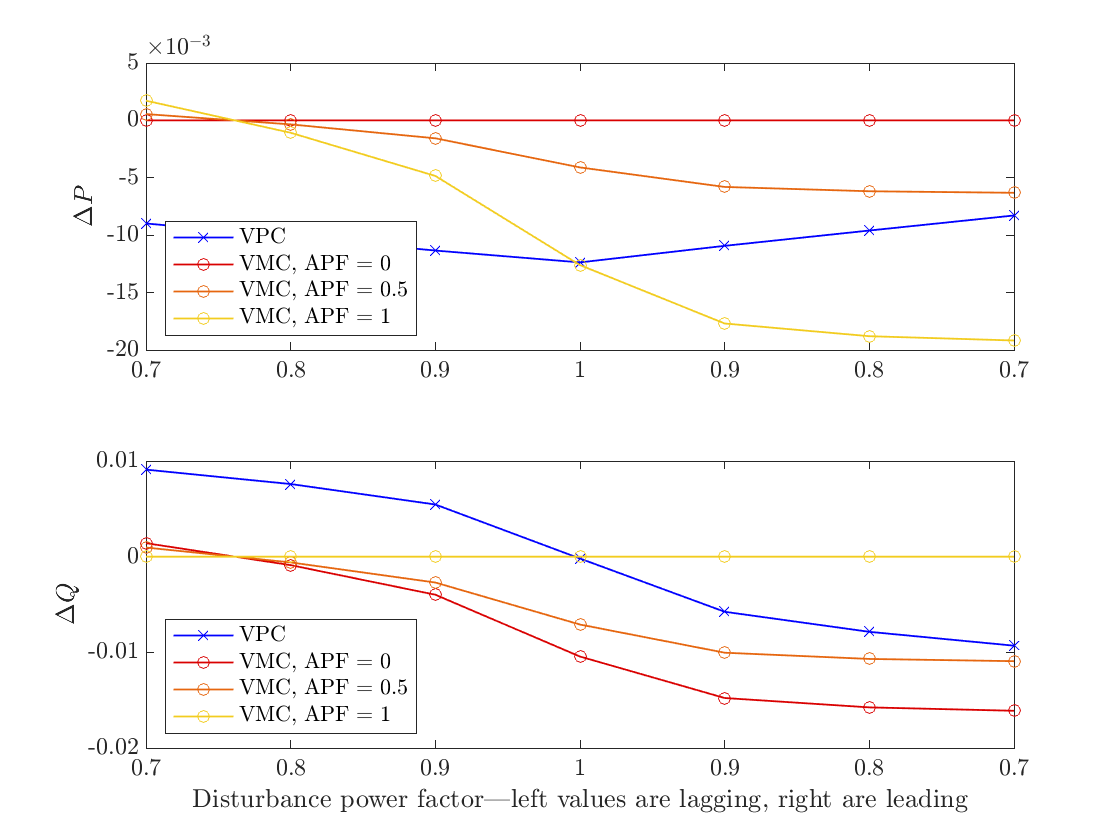}
\caption{\small The changes in node 2's power injections corresponding to Fig. \ref{fig:highVnonNominalLineflows}.}
\label{fig:highVnonNominalPQinj}
\end{figure}

The first set of simulations use a pre-disturbance baseline with excess distributed generation, similar to the scenarios considered in \cite{horowitz2020techno, procopiou2019limitations}, and \cite{jain2020dynamic}.
Both nodes 1 and 2 injected .06 W and extracted .02 VAR giving a .95 lagging power factor\footnote{We use the convention that a ``lagging'' power factor extracts VARs, regardless of whether real power is being injected or extracted.} and pre-disturbance $v_2 = 1.05\angle 1^\circ$.
The current magnitudes were normalized so that the pre-disturbance $|i_{01}| = 1$.
Disturbance injections with a range of power factors from 0.7 to 1 lagging, and from 1 to 0.7 leading were injected at node 1.
The PFC or MFC at node 2 adjusted its injection in order to maintain $\hat{v}_2 = 1.05\angle 1^\circ$ or $|v_2| = 1.05$, respectively. Fig. \ref{fig:highVnonNominalLineflows} plots $|i_{01}|$ vs. the power factor of the disturbance. 
The black line is $|i_{01}^\textnormal{bef}|$, the magenta line with squares is $|i_{01}^\textnormal{ol}|$, and the blue line with x markers is $|i_{01}^\textnormal{VPC}|$. The red, orange, and yellow lines with circles represent the post-disturbance $|i_{01}|$ with VMC applied to node 2 with lagging power factors of 0, 0.5, and 1, respectively.
Fig. \ref{fig:highVnonNominalPQinj} plots the real and reactive power injection adjustments (control effort) at node 2 that the PFC and MFCs used to maintain the assigned voltage targets. 

Fig. \ref{fig:highVnonNominalLineflows} demonstrates that VPC reduces the effect of the disturbance on $|i_{01}|$, regardless of the disturbance power factor. 
With APF = 0, VMC slightly increases the effect of the disturbance on $|i_{01}|$ for all but the most lagging power factor disturbance. This scenario is similar to the scenarios in \cite{procopiou2019limitations} and \cite{jain2020dynamic} in which Volt-VAR control exacerbated the line flow/thermal constraint violations because the Volt-VAR controllers' VAR extractions reduced the power factor of the line flows.
With APF = 0.5, VMC slightly reduces the effect of the disturbance on $|i_{01}|$ for leading power factor disturbances. 
With APF = 1, VMC reduces the effect of the disturbance on $|i_{01}|$ significantly for leading power factor disturbances. VMC's overreaction for leading power factor disturbances is explained by Fig. \ref{fig:highVnonNominalPQinj}---the APF $= 1$ VMC brings the voltage at node 2 back down by extracting a lot of real power, which is expensive. This real power extraction cancels some of the pre-disturbance baseline real power flow from node 0 to 1.



\ifdefined\fullresults

\subsection{Simulation set II: non-nominal baseline, low voltage example}
In the second set of simulations, the pre-disturbance baseline for the test was $v_2 = .95\angle 1^\circ$ and $|i_{01}| = 0.05$.
The current magnitudes were again normalized so that the pre-disturbance $|i_{01}| = 1$, the same disturbance injections were applied, and the PFC or MFC at node 2 adjusted its injection in order to maintain $\hat{v}_2 = .95\angle 1^\circ$ or $|v_2| = .95$, respectively. Fig. \ref{fig:nonNominalLineflows} demonstrates that, for the given non-nominal baseline and set of disturbance injections at node 1, both VPC and VMC decrease the effect of the disturbance on $|i_{01}|$, and that VPC generally does a better job. 
VMC does a good job of reducing the change in $|i_{01}|$ when the disturbance power factor is lagging because both the real and reactive portions of the disturbance injections decrease $|v_2|$. 
VMC does not do a good job of reducing the change in $|i_{01}|$ when the disturbance power factor is leading because both the reactive power injection counteracts the real power extraction, resulting in little change in $|v_2|$. 

\begin{figure}
\centering     
\includegraphics[scale=.22]{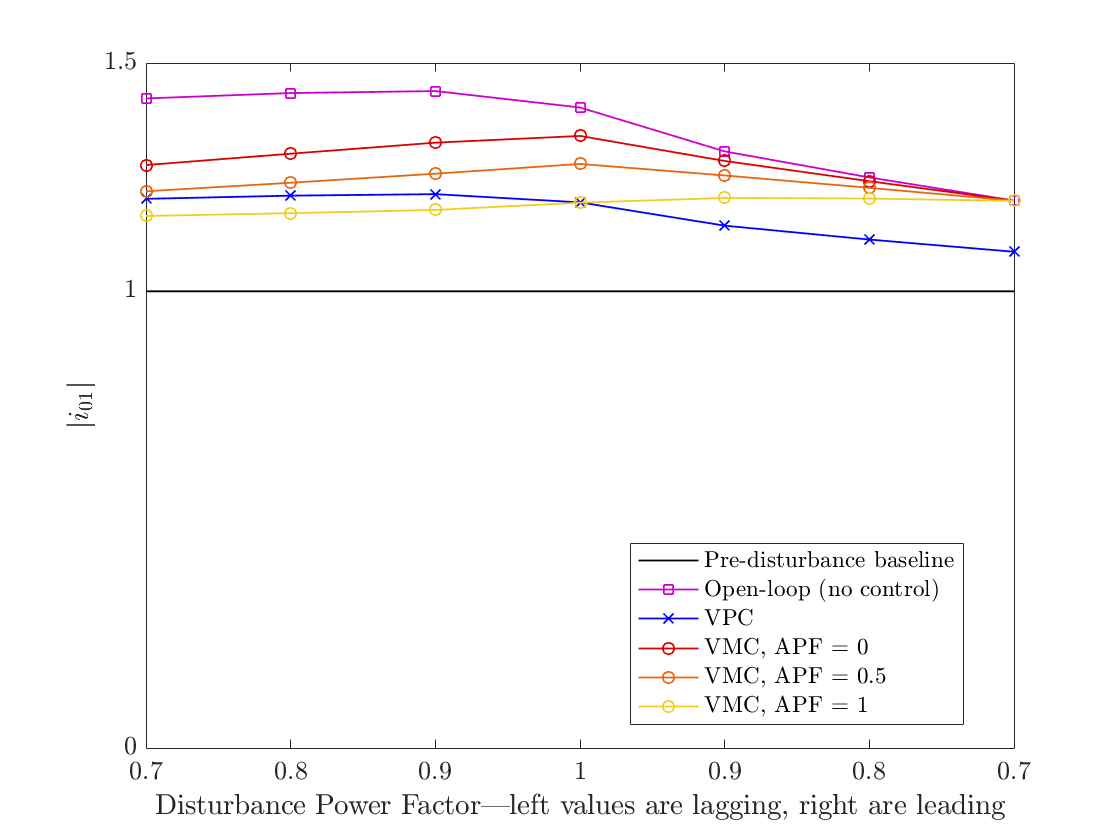}
\caption{\small Line flows for the circuit in row 3 of Table \ref{tab:Lfsens} with a non-nominal pre-disturbance baseline.}
\label{fig:nonNominalLineflows}
\end{figure}


\fi

\ifdefined\fullresults
\subsection{Simulation set III: nominal baseline}
\else
\subsection{Simulation set II: nominal baseline}
\fi 

\begin{figure}
\centering     
\includegraphics[scale=.22]{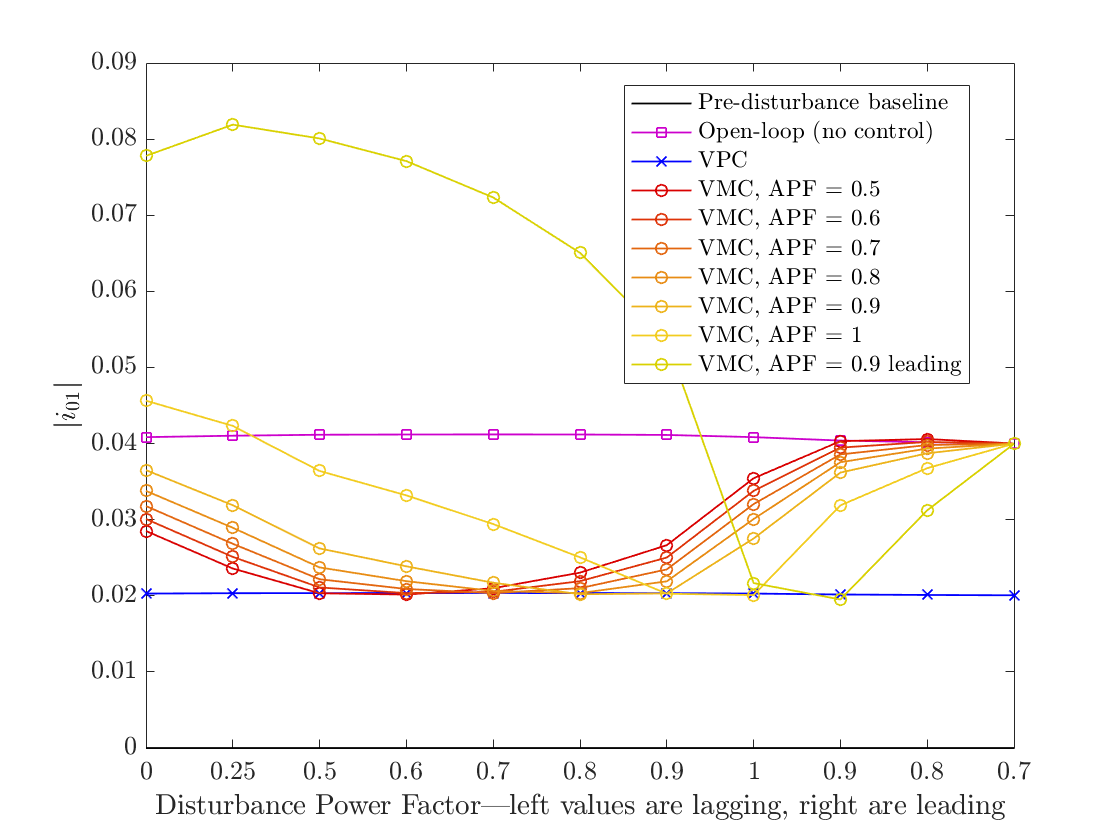}
\caption{\small Line flows for the circuit in Row 3 of Table \ref{tab:Lfsens} with the nominal pre-disturbance baseline.}
\label{fig:nominalLineflows}
\end{figure}

In the second set of simulations, the pre-disturbance baseline for the test was $i_1 = i_2 = 0$, which results in $v_1 = v_2 = 1$ and $|i_{01}| = 0$ (the ``nominal'' baseline).
Disturbance loads with a range of power factors from 0 to 1 lagging 
and from 1 to 0.7 leading were extracted from node 1, lowering node 2's voltage below 1.
The PFC or MFC at node 2 adjusted its injection in order to maintain $v_2 = 1$. 
In addition to the standard power factors, we also included the ``APF = 0.9 leading'' power factor VMC. 
Unlike the standard power factor VMCs, when the voltage magnitude is higher than the target, the APF = 0.9 leading VMC extracts real power and counter-intuitively \emph{injects} reactive power. When the voltage is lower than the target, it does the opposite.
Fig. \ref{fig:nominalLineflows} plots $|i_{01}|$ vs. the power factor of the disturbance and provides a number of insights.


First, the 0.7 leading power factor disturbance injection does not result in any change in the voltage magnitude at node 2. Thus, none of the VMCs adjust the power injection at node 2, and therefore do not reduce the increase in the upstream line flow. The VPC, on the other hand, recognizes the change in the voltage phasor angle at node 2, and adjusts its real and reactive power injections so that the increase in the line flow is one half of the no-control case. 

Second, the VMC can exacerbate the effect of the disturbance on $|i_{01}|$. This is demonstrated by the 0.9 leading APF for disturbances with lagging power factors, and by the APF = 1 Volt-Watt controller when the disturbance power factor is close to zero.

Third, 
Fig. \ref{fig:nominalLineflows} demonstrates that the VMC that reduces the effect of the disturbance on the upstream line flow the most is the VMC for which the APF is matched to the power factor of the disturbance. 

Fourth, the VPC injection is very similar to the VMC injection when the VMC's APF matches the disturbance injection's power factor. 
This is true even when the disturbance power factor is leading---when the disturbance power factor is leading and the voltage is low, the VPC \emph{extracts} reactive power and injects real power. Extracting reactive power when the voltage is low is non-intuitive, but is the correct response for reducing the impact of a leading power factor disturbance.
This VPC--VMC matching characteristic is not a hard rule for all pre-disturbance baselines and line impedances, however it does provide intuition for how VPC responds to disturbances: VPC injections tend to have power factors similar to the power factors of the disturbance injections that prompted the VPC injections.



\begin{insight}
VPC usually outperforms VMC in reducing the effect of disturbances on line flows that are upstream of both the disturbance and the controlled node in two circumstances:
\begin{enumerate}
    \item when the power factor of VMC does not match the power factor of the disturbance;
	\item when the disturbance has a leading power factor.
\end{enumerate}
\end{insight}



\section{Conclusion}

In this paper we introduced Voltage Phasor Control as a novel approach for implementing OPF which incorporates distributed feedback controllers. We described how VPC feedback rejects the effects of disturbance injections on the network's voltage magnitudes and upstream line flows. 
Through simulations, we compared the performance of VPC and VMC. 
These results establish VPC as a tool that can help distribution networks operate safely and reliably within their constraints. 
Ongoing research directions include alternative distributed feedback laws that incorporate the voltage angle measurement, as well as the VPC power flow optimization formulation.



\begin{appendices}
\section{Derivations of the voltage sensitivities in Table \ref{tab:Vsens}}\label{app:Vsens}


The sensitivities of the circuits without VPC are straightforward derivatives with respect to the disturbance injection current. 
With VPC, the PCN current injection at the PCN is a function of the disturbance node injection.

\subsection{Circuit in row 1 of Table \ref{tab:Vsens}}
\noindent The circuit equations are
\begin{gather*}
    v_1 = 1 - i_{01} z_{10}, \textnormal{ and}\\
    i_{01} = -(i_1 + i_2 + i_3).
\end{gather*}
\noindent Without VPC:
\begin{gather*}
    \frac{\partial v_1}{\partial i_3} = z_{10}.
\end{gather*}

\noindent With VPC:
\begin{gather}\label{eqn:V2PCN}
    i_2 = \frac{\hat{v}_2 - i_3(z_{10} + z_{21}) - i_1 z_{10} - 1}{z_{10} + z_{21}}.
\end{gather}
Therefore, $\frac{\partial i_2}{\partial i_3} = -1$. From KCL,
\begin{gather*}
    \frac{\partial i_{01}}{\partial i_3} = -\frac{\partial i_2}{\partial i_3} - 1 = 0.
\end{gather*}
Using the chain rule, we find that
\begin{gather*}
    \frac{\partial v_1}{i_3} = \frac{\partial v_1}{\partial i_{01}} \frac{\partial i_{01}}{i_3} = 0.
\end{gather*}

\subsection{Circuit in row 2 of Table \ref{tab:Vsens}}
\noindent The circuit equation is
\begin{gather}\label{row2CircEqn}
    v_3 = 1 + i_3 z_{32} + (i_2 + i_3)z_{21} + (i_3 + i_2 + i_1)z_{10}.
\end{gather}
\noindent Without VPC:
\begin{gather*}
    \frac{\partial v_3}{\partial i_1} = z_{10}.
\end{gather*}

\noindent With VPC, Eqn. (\ref{eqn:V2PCN}) still holds. Therefore, 
\begin{gather*}
    \frac{\partial i_2}{\partial i_1} = -\frac{z_{10}}{z_{10} + z_{21}},
\end{gather*}
and, from the circuit equation (\ref{row2CircEqn}),
\begin{align*}
    \frac{\partial v_3}{\partial i_1} &= \frac{\partial i_2}{\partial i_1} z_{21} + \frac{\partial i_2}{\partial i_1} z_{10} + z_{10}\\
    &= 0.
\end{align*}


\section{Derivations of the line flow sensitivities in Table \ref{tab:Lfsens}}\label{app:LFsens}


\noindent The circuit equations are
\begin{gather}
    1 - v_2 = i_{01} z_{10} + i_{12} z_{12}, \nonumber\\
    i_{01} = - i_1 + i_{12}, \textnormal{ and} \label{eqn:i01KCL}\\
    i_{12} = -i_2 - i_3. \label{eqn:i12KCL}
\end{gather}
When VPC holds $v_2$ at $\hat{v}_2$, $i_{01}$ and $i_{12}$ are functions of $i_1$:
\begin{gather}
    i_{01} = \frac{1 - \hat{v}_2}{z_{10} + z_{12}} - i_1 \frac{z_{12}}{z_{10} + z_{12}}. \label{eqn:i01PCN}\\
    i_{12} = \frac{1 - \hat{v}_2}{z_{10} + z_{12}} + i_1 \frac{z_{01}}{z_{10} + z_{12}}, \label{eqn:i12PCN}
\end{gather}
The sensitivities in Table \ref{tab:Lfsens} come from the partial derivatives of equations (\ref{eqn:i01KCL}), (\ref{eqn:i12KCL}), (\ref{eqn:i01PCN}), and (\ref{eqn:i12PCN}).









\section{Proof of Lemma \ref{lemma:DeltaCurr01}}\label{app:LemmaDeltaCurr}
\begin{proof}

From (\ref{eqn:i01PCN}), 
$\frac{\partial i_{01}}{\partial i_1} = -a$.\\
Using the Triangle Inequality,
$$
|i^\textnormal{VPC}_{01}| - |i^\textnormal{bef}_{01}| \leq |\Delta i_1 \frac{\partial i_{01}}{\partial i_1}| = | \Delta i_1||a|.
$$
\end{proof}
\noindent A similar proof applies to Lemma \ref{lemma:DeltaCurr12}.

\section{Proof of Lemma \ref{lemma:CurrDiff}}\label{app:Lemma:CurrDiff}
\begin{proof}

    Without VPC, $i_{01}^\textnormal{ol} = (i^\textnormal{bef}_{01} - \Delta i_1)$, and
    \begin{align*}
        |i_{01}^\textnormal{ol}|^2 &= (i^\textnormal{bef}_{01} - \Delta i_1)(i^\textnormal{bef}_{01} - \Delta i_1)^*\\
        &= |i^\textnormal{bef}_{01}|^2 - 2\cos(\theta_{i^\textnormal{bef}_{01}} - \theta_{\Delta i_1})|i^\textnormal{bef}_{01}||\Delta i_1| + |\Delta i_1|^2.
    \end{align*}    
    With VPC, $i_{01}^\textnormal{VPC} = (i^\textnormal{bef}_{01} - a \Delta i_1)$, and
    \begin{align*}
        |i_{01}^\textnormal{VPC}|^2 &= (i^\textnormal{bef}_{01} - a \Delta i_1)(i^\textnormal{bef}_{01} - a \Delta i_1)^*\\
        &= |i^\textnormal{bef}_{01}|^2 - 2a \cos(\theta_{i^\textnormal{bef}_{01}} - \theta_{\Delta i_1})|i^\textnormal{bef}_{01}||\Delta i_1| + a^2|\Delta i_1|^2.
    \end{align*}    
    Because $\Delta i_1$ increases $i_{01}$ and $z_{01}$ and $z_{12}$ have the same X/R ratio, $|i_{01}^\textnormal{VPC}|^2 > |i^\textnormal{bef}_{01}|^2$, and 
    \begin{align*}
        0 < -2a \cos(\theta_{i^\textnormal{bef}_{01}} - \theta_{\Delta i_1})|i^\textnormal{bef}_{01}||\Delta i_1| + a^2|\Delta i_1|^2.
    \end{align*}
    Using the fact that $0 < a < 1$, 
    \begin{align}\label{eqn:lemmaUsefulBound1}
        0 < 2(a -1) \cos(\theta_{i^\textnormal{bef}_{01}} - \theta_{\Delta i_1})|i^\textnormal{bef}_{01}||\Delta i_1| + (a - a^2)|\Delta i_1|^2.
    \end{align}
    Taking the difference between $|i_{01}^\textnormal{ol}|^2$ and $|i_{01}^\textnormal{VPC}|^2$, we get
    \begin{align}
        |i_{01}^\textnormal{ol}|^2 - |i_{01}^\textnormal{VPC}|^2 &= 2(a - 1)\cos(\theta_{i^\textnormal{bef}_{01}} - \theta_{\Delta i_1})|i^\textnormal{bef}_{01}||\Delta i_1| \nonumber\\&+ (a - a^2)|\Delta i_1|^2 + (1 - a)|\Delta i_1|^2. \label{eqn:lemmaDiff}
    \end{align}    
    Substituting (\ref{eqn:lemmaUsefulBound1}) into (\ref{eqn:lemmaDiff}) gives
    \begin{align*}
        |i_{01}^\textnormal{ol}|^2 - |i_{01}^\textnormal{VPC}|^2 > (1 - a)|\Delta i_1|^2, 
    \end{align*}
    which is positive because $0 < a < 1$.
\end{proof}

\end{appendices}



%



\bibliographystyle{IEEEtran}
\bibliography{references.bib}

\end{document}